\documentclass[a4paper,11pt]{article}
\usepackage{jheppub}

\pdfoutput=1
\usepackage{hyperref}
\usepackage{graphicx}
\usepackage[utf8]{inputenc}
\usepackage{xspace}
\usepackage{amsmath,amssymb}
\usepackage{bbm}
\usepackage{xcolor}
\usepackage{booktabs}
\usepackage{subcaption}
\usepackage{mathtools}
\usepackage[ruled,norelsize]{algorithm2e}
\usepackage{empheq}
\usepackage{cancel}
\usepackage{multirow}
\usepackage{slashed}
\usepackage{bbold}
\usepackage{cleveref}
\usepackage{algorithm2e}

\DeclareFontFamily{U}{rcjhbltx}{}
\DeclareFontShape{U}{rcjhbltx}{m}{n}{<->rcjhbltx}{}

\title{Chebyshev Approximations of Feynman Integrals for Collider Physics}

\author[a,b]{Samuel Abreu,}
\author[c,d]{Afonso Guerreiro,}
\author[e]{Ben Page}

\affiliation[a]{CERN, Theoretical Physics Department, CH-1211 Geneva
  23, Switzerland}
\affiliation[b]{Higgs Centre for Theoretical Physics, School of Physics and Astronomy,
 The University of Edinburgh, Edinburgh EH9 3FD, Scotland, United Kingdom}
 \affiliation[c]{
 Instituto Superior Técnico (IST), Universidade de Lisboa,
Av. Rovisco Pais 1, P-1049-001 Lisboa, Portugal
 }
 \affiliation[d]{
 LIP, Av. Prof. Gama Pinto, 2, P-1649-003 Lisboa, Portugal
 }
\affiliation[e]{Department of Physics and Astronomy, Ghent University, 9000 Ghent, Belgium}

\emailAdd{samuel.abreu@cern.ch}
\emailAdd{afonsojguerreiro@tecnico.ulisboa.pt}
\emailAdd{ben.page@ugent.be}

\preprint{CERN-TH-2026-160}
\abstract{
We present a novel approach for solving canonical differential equations for Feynman integrals based on an approximation of the integrals with Chebyshev polynomials. 
By exploiting the analyticity properties of Feynman integrals, the method constructs rapidly converging polynomial approximations along a path, enabling highly efficient numerical evaluation. 
Moreover, we introduce an adaptive approximation method that dynamically samples to optimise convergence. 
We implement this framework in double-precision arithmetic and demonstrate its stability across physical phase space using a series of two-loop, five-point examples. 
Our proof-of-principle implementation proves competitive with state-of-the-art one-fold integral methods, while requiring little to no case-by-case intervention to handle spurious singularities. 
}

\keywords{}
\begin{document}
\setlength{\parskip}{0pt}
\maketitle
\flushbottom

\section{Introduction}
\label{sec:intro}

Feynman integrals are fundamental building blocks required to carry out computations in Quantum Field Theory. 
Their evaluation is essential for precisely probing the Standard Model and searching for new physics. 
In the context of high-energy collision experiments~\cite{Collider_Physics_Precision_Frontier,Les_Houches_2023,Anomalous_Magnetic_Moment}, 
the increase in experimental precision demands the inclusion of higher-order terms in perturbation theory, thereby necessitating the calculation of multi-loop Feynman integrals. The systematic evaluation of such integrals, particularly in the presence of multiple kinematic scales, remains an active area of research.
Currently, the state-of-the-art approach consists of reducing all Feynman integrals arising in a given calculation to a smaller set of master integrals via systematic use of integration-by-parts (IBP) identities~\cite{IBP}. 
The solution of the IBP identities is obtained with Laporta's algorithm~\cite{Laporta}, 
which has been implemented in many public codes~\cite{KIRAV3,Smirnov:2025prc,Reduze2,
Guan:2024byi,% Blade
Lee:2013mka,% LiteRed
Wu:2025aeg% NeatIBP
}. 

While there exist many techniques for the evaluation of master integrals (see, e.g. \cite{Smirnov_Feynman_Calculus,Stefan_Weinzierl_Feynman_Integrals} for detailed reviews), the most prominent among them is the method of differential equations \cite{Kotikov:1990kg,Kotikov:1991pm,
Bern:1993kr,Remiddi:1997ny,Gehrmann:1999as}. 
Here, one differentiates the master integrals with respect to kinematical variables and internal masses, leading to a system of first-order differential equations whose solution yields the master integrals. 
In many cases, this system can be cast into so-called ``canonical form''~\cite{Henn:2013pwa} in which the dependence on the dimensional regulator, $\epsilon$, factorises, and the integration kernels are $\mathrm{d}\log$ forms. 
In this form, the differential equations admit a straightforward solution in terms of iterated integrals, which can often be expressed in terms of multiple polylogarithms. 
Furthermore, for multi-scale problems, even when such a basis can be found, finding an analytic solution is often prohibitively difficult and one must often reach for numerical methods.

To address this issue, several numerical methods have been developed for solving the differential equations satisfied by the master integrals. 
One such method relies on power-series expansions to construct local solutions~\cite{Generalized_Power_Series,Moriello:2019yhu}. 
This technique has been implemented in the publicly available \texttt{Mathematica} packages \textsc{DiffExp} \cite{Hidding:2020ytt} and \textsc{Seaside} \cite{Armadillo:2022ugh}. A similar idea underpins the auxiliary mass flow method \cite{Auxiliary_Mass_Flow_1,Auxiliary_Mass_Flow_2}, 
which has also been implemented in the \texttt{Mathematica} package \textsc{AMFlow} \cite{Liu:2022chg}. 
These implementations are widely used to provide boundary conditions and evolve a solution over phase-space. However, despite this success, they often exhibit long runtimes, making them challenging to adopt for phenomenological applications. 
The package \textsc{LINE} \cite{Prisco:2025wqs} seeks to mitigate these performance limitations by adopting a \textsc{C++} implementation of the auxiliary mass flow method. 
Moreover, recent investigations into novel techniques for numerically solving differential equations such as Runge-Kutta and Bulirsch-Stoer integration~\cite{Badger:2025ljy,Badger:2025ilt,PetitRosas:2025xhm,Czakon:2026tog} have shown impressive runtime for frontier two-loop examples, and
analogous approaches are being explored to numerically compute iterated integrals~\cite{Baur:2026zlw}. 
More recently, pseudo-spectral methods have been proposed and demonstrated for the evaluation of two-loop massless six-point integrals \cite{Liu:2026hdp}, resulting in highly accurate numerical evaluations.

In this work, we propose an alternative approach based on global expansions in terms of Chebyshev polynomials to approximate a given target function.
This sets the method apart from local expansion 
schemes, such as those based upon generalised power series expansions. 
Since first being introduced in the context of numerical fluid dynamics 
(see \cite{Spectral_Methods_Fluid_Mechanics_1,Spectral_Methods_Fluid_Mechanics_2}), 
they have been applied in numerous contexts spanning from meteorology and 
climate modelling \cite{Metereology} to numerical general relativity \cite{General_Relativity}.
Their increasing popularity can be largely attributed to their excellent convergence properties: 
for analytic functions, their convergence is exponential.
This feature motivates our study of their use in applications to Feynman integrals, 
whose region of analyticity is becoming increasingly well understood in recent years~\cite{Helmer:2025ljj,Mizera:2021icv,Fevola:2023kaw,Fevola:2023fzn,Correia:2025wtb,Helmer:2024wax,Dlapa:2023cvx}.

In this work, we exploit the suitability of Chebyshev polynomials to approximate analytic 
functions in order to construct an efficient method for numerically evaluating
multi-scale Feynman integrals over physical phase space. 
Our approach leverages modern understanding of numerical construction of Chebyshev approximations.
Specifically, we use tools introduced in the \texttt{chebfun} project~\cite{Chebfun,chebfun1,Aurentz2017} to construct an adaptive Chebyshev 
approximation method that avoids oversampling in the presence of roundoff error and nearby non-analyticities, optimizing convergence.
We study our approach in applications to two-loop five-point Feynman integrals, stress testing our 
algorithm by using it to evaluate so-called ``pentagon functions'' for full-colour five-point massless matrix elements and 
leading-colour five-point one-mass matrix elements.
In practice, we find that the approach is numerically stable, and is insensitive to the presence of spurious 
singularities of the differential equation. Comparing it to cutting-edge implementations of one-fold integration techniques~\cite{Chicherin:2020oor,Chicherin:2021dyp}, we find competitive runtimes and stability properties.
We provide a proof of principle \texttt{Mathematica} implementation in ancillary files to enable future study.

This work is structured as follows. 
In \cref{sec:CanonicalDEs}, we introduce the necessary details of canonical differential equations for Feynman integrals. 
In \cref{sec:spectralmethods}, we describe our Chebyshev-based solution method. We begin by discussing 
the basic theory of Chebyshev polynomials and Chebyshev interpolants as well as their convergence 
properties in numerical applications. We then discuss our adaptive Chebyshev approximation approach.
In \cref{sec:pentagonFunctions}, we apply our approach to a collection of two-loop five-point Feynman 
integrals, and study their numerical stability over phase space, providing a proof-of-concept approach
in ancillary files.

\section{Canonical Differential Equations for Feynman Integrals}
\label{sec:CanonicalDEs}

In this work, we are interested in studying the application of Chebyshev polynomial approximation
methods to solve the differential
equations that Feynman integrals
satisfy~\cite{Kotikov:1990kg,Kotikov:1991pm,
Bern:1993kr,Remiddi:1997ny,Gehrmann:1999as,Henn:2013pwa}.
In recent years, the differential equations method has become one of the most
commonly used approaches for computing Feynman integrals---prominently in the case
of multi-scale Feynman integrals.
Denoting a basis of master integrals as $\vec{J}$,
the integrals are functions of the kinematic invariants $\vec{s}$ and
the dimensional regulator $\epsilon = (4-D)/2$.
Such a basis of master integrals satisfies a first-order partial
differential equation as
\begin{equation}
 \mathrm{d} J_i = A_{ij}(\vec{s},\epsilon) J_j,
 \label{eq:GeneralDE}
\end{equation}
where the $A(\vec{s}, \epsilon)$ are matrices of one forms that are
rational in the invariants $\vec{s}$ and $\epsilon$.
If one has such a system of differential equations to hand, the problem of
computing the collection of Feynman integrals is reduced to evaluating an
appropriate boundary condition and solving the differential equation to evolve
this boundary condition to another point in the physical phase space.

The difficulty of solving the differential equation can be alleviated by an appropriate
choice of basis. If we consider the differential equation for a different basis of Feynman integrals 
$\vec{I}$ related to $\vec{J}$ via $J_i = U_{ij} I_j$, then the new basis satisfies the differential equation
\begin{equation}
    \mathrm{d} I_i = \tilde{A}_{ij} I_j, \qquad \text{where} \quad \tilde{A} = U^{-1} A U - U^{-1} \mathrm{d} U.
    \label{eq:DEInNewBasis}
\end{equation}
In practice, solving the differential equation is arguably the simplest if one works with a basis of
master integrals that satisfies a differential equation that is in so-called
``canonical form''~\cite{Henn:2013pwa}.
If $\vec{I}$ is such a basis, then \cref{eq:DEInNewBasis} simplifies to the form
\begin{equation}
    \mathrm{d} I_i = \epsilon M_{ijk} \mathrm{d}\log(W_j) I_k,
    \label{eq:CanonicalDE}
\end{equation}
where $\epsilon = (4-D)/2$ is the dimensional regulator,
the coefficients $M_{ijk}$ are rational numbers, and the arguments $W_j$ are
algebraic functions of the kinematics. 
The analytic structure of
these Feynman integrals is strictly controlled by the collection of $W_j$, which
are typically referred to as the ``letters'' of the alphabet.
Not all collections of Feynman integrals admit a basis that satisfies a
canonical differential equation as defined above, for example in the presence of elliptic
integrals.
While it would be interesting to study the application of our approach to such
Feynman integrals, we leave this to future work.

For phenomenological applications of Feynman integrals in collider physics, the
principal aim is to evaluate the integrals as a series expansion in $\epsilon$.
This task is greatly facilitated by the $\epsilon$-factorised form of the
canonical differential equation in \cref{eq:CanonicalDE} as one can
write the series expansion of the vector of integrals as
\begin{equation}
    I_i(\vec{s}, \epsilon) = \sum_{n = 0}^\infty \epsilon^n I^{(n)}_i(\vec{s}),
    \label{eq:IntegralSeriesExpansion}
\end{equation}
where the $\epsilon$ expansion order $n$ denotes the weight of the integral. 
By inserting \cref{eq:IntegralSeriesExpansion} into \cref{eq:CanonicalDE}, one concludes that the coefficients of the $\epsilon$ expansion satisfy an iterative tower of differential equations given by
\begin{equation}
    \mathrm{d} I_i^{(n)} = M_{ijk} \mathrm{d}\log(W_j) I^{(n-1)}_k.
\end{equation}
Importantly, while the integrals $I_i$
are linearly independent to all orders in $\epsilon$, the individual coefficients $I_i^{(n)}$ can exhibit
linear dependencies.
It is therefore natural to consider solving for an independent set of functions that span the space at each weight, which 
we will denote as $\tilde{I}_i^{(n)}$.
In ref.~\cite{Abreu:2023rco}, it was proposed that linear dependencies between functions can be
easily resolved by working at the symbol level, supplemented by a small number of constraints from numerical evaluations.
This procedure generates both the basis $\tilde{I}_i^{(n)}$ and a rational
number matrix $A^{(n)}_{ij}$ such that the original coefficients map to the basis via
$I_i^{(n)} = A^{(n)}_{ij} \tilde{I}_{j}^{(n)}$.
Using this information, it is systematic to construct a differential equation for the 
independent special functions of the form
\begin{equation}
    \mathrm{d} \tilde{I}^{(n)}_i = \tilde{M}_{ijk}^{(n)} \mathrm{d}\log(W_j) \tilde{I}_k^{(n-1)},
    \label{eq:PentagonFunctionDEs}
\end{equation}
where the $\tilde{M}_{ijk}^{(n)}$ are rational numbers and $n \ge 1$.

The differential equation in \cref{eq:PentagonFunctionDEs} is particularly 
appealing as its formal solution along a path can easily be achieved by direct integration.
Let us consider a path in Mandelstam space $\vec{s}(t)$, which starts at $t = 0$ and ends at $t = 1$.
If we assume the availability of a boundary value $I^{(n)}(\vec{s}_0)$, one can then directly write
\begin{equation}
    \tilde{I}^{(n)}_i(\vec{s}[t]) = \tilde{I}^{(n)}_i(\vec{s}_0) + \int_0^t \mathrm{d}t' M_{ijk}^{(n)} \frac{\partial}{\partial t'} \log(W_j[\vec{s}(t')]) \tilde{I}_k^{(n-1)} (\vec{s}[t']).
    \label{eq:PathIntegrationForm}
\end{equation}
The challenge in applying \cref{eq:PathIntegrationForm} is to perform the 
integration on the right-hand side. If performed analytically, this integration leads 
to the introduction of dedicated classes of special functions such as generalised 
polylogarithms.
Alternatively, if one can find an analytic representation for the integrands in \cref{eq:PathIntegrationForm}, one can compute these integrals numerically, as done in \cite{Chicherin:2020oor, Chicherin:2021dyp, Abreu:2023rco}.
In this work, we take yet another alternative, quasi-numerical, route. We will explore computing the integrals in 
\cref{eq:PathIntegrationForm} by using the differential equations to approximate the $\tilde{I}^{(n)}_i(\vec{s}[t])$ 
using Chebyshev polynomials. 

\section{Chebyshev Approximations of Feynman Integrals}
\label{sec:spectralmethods}

In recent years, a popular method that has arisen for solving the differential equations that Feynman integrals 
satisfy, \cref{eq:GeneralDE},  is that of ``generalised power series expansion''~\cite{Generalized_Power_Series,Moriello:2019yhu,Hidding:2020ytt,Prisco:2025wqs,Armadillo:2022ugh, Liu:2022chg}. Here, one begins by choosing a univariate path $\vec{s}(t)$ in phase space 
between an initial point $\vec{s}_0$ and a final point $\vec{s}_1$ where the value of the set of Feynman integrals 
is desired. The Feynman integrals are then locally represented as generalised power series in the path parameter $t$, constrained 
to satisfy the differential equation.
This allows both for a systematic evaluation of the set of Feynman integrals at the point $\vec{s}_1$, as well as a 
quasi-analytic solution for the Feynman integral everywhere along the path.

One of the practical difficulties of this approach is that singularities of the differential equation which live in the complex $t$-plane 
can limit the radius of convergence of a generalised power series centred at $\vec{s}_0$ so that it does not include the point $\vec{s}_1$.
The common solution to this problem is to employ a collection of power-series solutions, each with an overlapping region of convergence, 
which continuously cover the integration path and thereby evolve the solution from $\vec{s}_0$ to $\vec{s}_1$. 
Nevertheless, from both a practical and physical perspective, this feature is undesirable. Specifically, when evolving over paths 
contained within physical phase-space, singularities that lie directly along the path are very rarely encountered (see e.g. ref.~\cite{Abreu:2023rco} for an example arising in non-planar Feynman integrals). 
Moreover, in recent years, our 
understanding of the location of singularities of Feynman integrals has become increasingly systematic~\cite{Helmer:2025ljj,Mizera:2021icv,Fevola:2023kaw,Fevola:2023fzn,Correia:2025wtb,Helmer:2024wax,Dlapa:2023cvx}. It is therefore 
natural to ask if there exists a method that is similarly powerful to power series expansion, but is (in principle) only sensitive to 
singularities that lie along the integration path, not in the complex $t$-plane. In this section, we discuss how a representation 
of Feynman integrals as a ``Chebyshev series'' can answer this question. 
We will begin by introducing the necessary theoretical background on Chebyshev series, then discuss how they can be manipulated, and finally 
discuss how they can be constructed in a practical manner that ensures numerical stability.

\subsection{Chebyshev Series and Interpolants}
\label{sec:Chebysev_Series}

The starting point for our discussion is the following important result~\cite{Lloyd_Approximation_Theory}: any sufficiently 
continuous\footnote{The details of the continuity assumption here are somewhat technical. 
Lipschitz continuity is sufficient for absolute convergence, while for pointwise convergence continuity is sufficient. 
Furthermore, for uniform convergence, a sufficient condition is that the function should have bounded variation or 
satisfy the Dini-Lipschitz condition \cite{Chebyshev_Review}. } function $f(x)$ defined on the interval $[-1, 1]$ admits a unique
representation as a ``Chebyshev series''.
Specifically, one can write:
\begin{equation}
    f(x) = \sum_{k=0}^\infty a_k T_k(x), \qquad a_k = \frac{2}{(1+\delta_{k 0})\pi}\int_{-1}^1\frac{f(x)T_k(x)}{\sqrt{1-x^2}} \mathrm{d} x,
    \label{eq:Chebysev_Series}
\end{equation}
where the $T_k(x)$ are polynomials of degree $k$ in $x$ known as the ``Chebyshev polynomials of the first kind''.
For ease we will refer to the $T_k(x)$ simply as Chebyshev polynomials.
Such Chebyshev polynomials can be defined implicitly through the relation
\begin{equation}
    T_k(\cos{\theta}) = \cos{k\theta},
    \label{eq:Chebyshev_Trignometric_Argument}
\end{equation}
where standard trigonometric relations make it clear that the right hand side is a polynomial in $\cos(\theta)$.
A comprehensive discussion of Chebyshev polynomials can be found in ref.~\cite{Chebyshev_Review}. 
Note that, even though \cref{eq:Chebysev_Series} applies to the interval $[-1,1]$, we may extend its application to any finite
interval $[a,b]$ via some function that maps $[a,b] \leftrightarrow [-1,1]$. (We will return to the question of choosing a suitable map in \cref{sec:AdaptiveAlgorithmDescription}.)

To consider the convergence properties of Chebyshev series in more detail, let us introduce the concept of a ``Bernstein ellipse'' \cite{BERNSTEIN_ELLIPSE_1,
BERNSTEIN_ELLIPSE_2}. This is defined as the image of a complex
circle $|z| = \rho$ which is centred at the origin, under the ``Joukowsky map''
\begin{equation}
    z  \mapsto \frac{z+z^{-1}}{2}.
\end{equation}
For $\rho > 1$, this mapping yields an ellipse in the complex plane with foci at
$ z= \pm 1$. Its major (minor) semi-axis $a$ ($b$) is given in terms of $\rho$ by
\begin{equation}
    a = \frac{1}{2}(\rho+\rho^{-1}), \quad \quad b= \frac{1}{2}(\rho-\rho^{-1}).
\end{equation}
Now, let $E_\rho$ denote the interior of the largest Bernstein ellipse to which $f$ can be extended analytically.
For our purposes, the importance of this Bernstein ellipse 
 is that the Chebyshev series of $f$ converges inside $E_\rho$ and diverges everywhere
outside it \cite{Chebyshev_Boyd}. 

The parameter $\rho$ is determined by the location of the nearest singularity of
$f$ to the interval $[-1,1]$ in the complex plane, i.e. the closest point beyond
which $f$ can no longer be analytically continued. If the convergence-limiting
singularity is located at $z_0 = x_0 + i y_0$ in the complex plane, then the
parameter $\rho$ for the largest admissible Bernstein ellipse is given by:
\begin{equation}
    \rho = \alpha +\sqrt{\alpha^2-1}, \quad \alpha = \frac{1}{2}\left[\sqrt{(x_0+1)^2+y_0^2}+\sqrt{(x_0-1)^2+y_0^2}\right]
    \label{eq:Bernstein_Ellispe_Parameter}
\end{equation}
Note that the interval $[-1,1]$  is always contained within  $E_\rho$. As such, any singularity of $f(x)$ that lies 
outside $[-1,1]$ may slow the convergence of the series, but it can never spoil it.
For our applications to Feynman integrals, the importance of this statement is that if a Feynman integral is sufficiently continuous along a path, then it can be represented by a single Chebyshev series.

\begin{figure}[tb]
    \centering
    \includegraphics[scale=0.65]{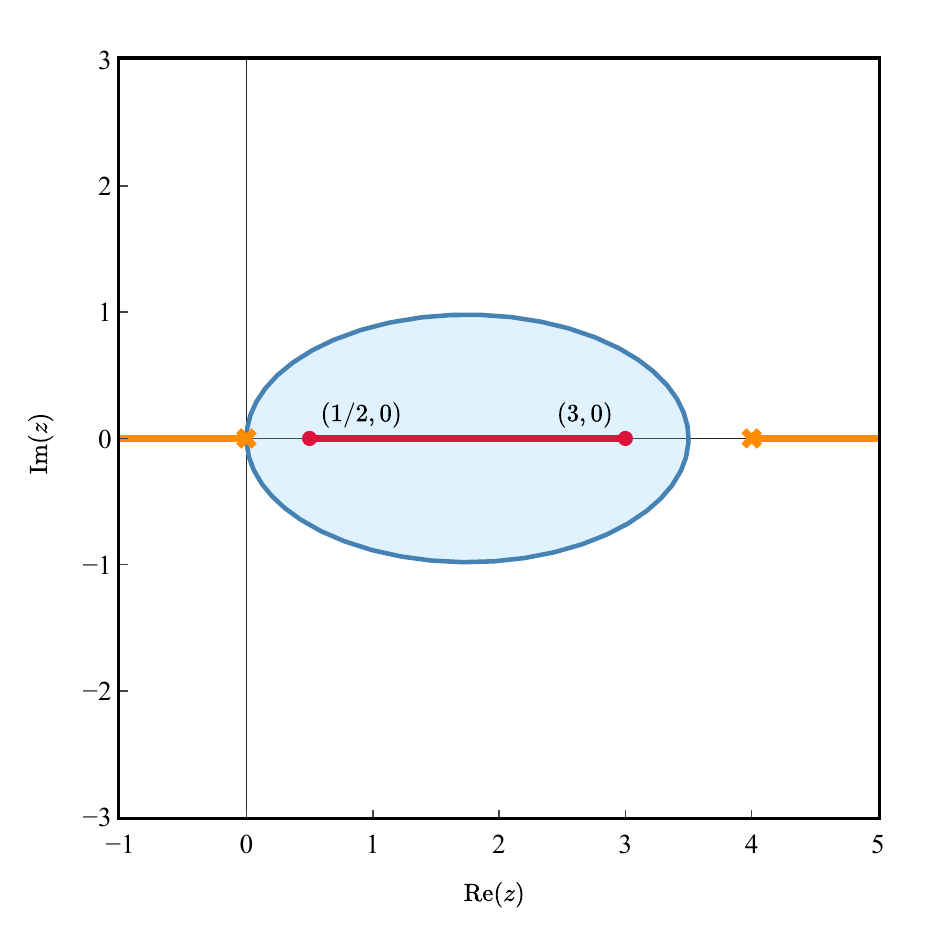}
    \caption{Region of convergence for the Chebyshev series representation of the pure 2-dimensional massive bubble integral $I_{\text{bub}}(s, m^2 = 1)$. The crimson dots at $(1/2,0)$ and $(3,0)$ indicate the foci of the ellipse, while the orange crosses (lines) represent the branch points (cuts) of $I(s,1)$.}
    \label{fig:Bernstein_Ellipse}
\end{figure}

For clarity, let us illustrate this discussion with a physical example. Specifically, we consider the pure equal mass one-loop bubble in 2 dimensions, 
\begin{equation}
    I_{\text{bub}}(s,m^2) = \log\left(\frac{\sqrt{4m^2-s}-i\sqrt{s}}{\sqrt{4m^2-s}+i\sqrt{s}}\right)\,,
    \label{eq:Bubble_Integral}
\end{equation}
which has branch points at $s=4m^2$ and $s=0$ with branch cuts for $s\in[4m^2,\infty)$ and $s\in (-\infty,0]$. 
Naturally, $I_{\text{bub}}(s, m^2 = 1)$ is not analytic for $s \in [-1, 1]$. However, we can easily map $x \in [-1, 1]$ to some other interval $[a,b]$ through
\begin{equation}
    s(x) = b - (b - a) \frac{1 - x}{2}.
\end{equation}
For concreteness, we choose the interval $[1/2,3]$, where $I_{\text{bub}}(s, m^2 = 1)$ is analytic. 
The region of convergence of the Chebyshev series is determined by the largest (mapped) Bernstein ellipse whose interior contains no singularities of $I_{\text{bub}}(s, 1)$. 
The closest branch point to the interval $[1/2,3]$ is located at $s=0$; therefore, the parameter corresponding to the largest admissible Bernstein ellipse is $\rho \approx 2.380$. The region of convergence of the Chebyshev interpolation of $I_{\text{bub}}(s, m^2 = 1)$ is illustrated in \cref{fig:Bernstein_Ellipse}.

Naturally, in computational applications, one cannot work directly with the infinite expansion in \cref{eq:Chebysev_Series}. 
Rather, we must construct a finite sum of Chebyshev polynomials. 
One approach to do this would be to truncate the series, keeping only the first $n+1$ terms.
However, the difficulty of computing the coefficients $a_k$ via the weighted integral in \cref{eq:Chebysev_Series}
means that this approach is rarely used in practice.
Alternatively, we can use interpolation. Specifically, we will construct a Chebyshev series that agrees with the function $f$ 
at a finite set of $n+1$ so-called Chebyshev nodes.\footnote{
Importantly, Chebyshev grids often outperform alternative node distributions such as equidistant grids. 
A textbook example is provided by Runge's phenomenon, in which polynomial interpolation on equidistant nodes exhibits increasingly large oscillations near the endpoints of the interval $[-1,1]$, leading to a catastrophic loss of accuracy. By contrast, interpolation on Chebyshev nodes is immune to this instability, converging uniformly on $[-1,1]$.
}
We will work with the so-called ``Chebyshev nodes of the second kind'' or ``Chebyshev-Lobatto nodes''. These
are defined to be the $n+1$ extrema of the $n$-th Chebyshev polynomial, and are
located at
\begin{equation}
    x_j = \cos\left(\frac{j}{n}\pi\right),\, \quad j\in\{0,...,n\}.
    \label{eq:Chebyshev_Extrema}
\end{equation}
We define a Chebyshev interpolant of $f(x)$ of order $n$ to be the unique polynomial $p_n$ of degree $n$ 
which matches $f(x)$ on $n+1$ Chebyshev-Lobatto nodes. That is,
\begin{equation}
    p_n(x_k) = f(x_k).
    \label{eq:ChebyshevLobattoNodeMatching}
\end{equation}
While such a polynomial can be represented using standard interpolation techniques in terms of 
so-called Lagrange cardinal polynomials~\cite{Lloyd_Approximation_Theory}, it is more natural to express it in terms of the Chebyshev basis as
\begin{equation}
    p_n(x) = \sum_{k=0}^n \hat{a}_k T_k(x). 
    \label{eq:ChebyshevInterpolantForm}
\end{equation}
Importantly, the coefficients of this Chebyshev expansion, $\hat{a}_k$, can be computed efficiently as we will discuss in \cref{sec:ChebyshevOperations}. 
We note that interpolation generally yields a different polynomial approximation from that obtained via truncation. 
That is, for a given $f(x)$, $a_k \ne \hat{a}_k$ in eqs~\eqref{eq:Chebysev_Series} and \eqref{eq:ChebyshevInterpolantForm}. 
Nevertheless, both constructions share similar convergence properties while typically yielding approximations 
of comparable accuracy and we thus choose to focus on interpolation for the remainder of this work.

In practice, the convergence of a Chebyshev interpolant is intimately related to the smoothness of the 
function being approximated. As a general rule of thumb, the smoother the function, the 
faster the rate of convergence. This observation can be made precise through the following 
theorems (for further details, we refer the reader to ref.~\cite{Lloyd_Approximation_Theory}), 
beginning with a result concerning differentiable functions. 
 Let $m\in \mathbb{N}_0$, 
and suppose that $f(x)$ and its derivatives up to $f^{(m-1)}$ are absolutely continuous on $[-1, 1]$, 
while the $m$-th derivative has finite variation\footnote{
 Letting $f$ be differentiable with a Riemann-integrable first derivative, one defines the 
``variation of $f$'' on an interval $[a,b]$ as
$V = \int_{a}^b |f'(x)|dx\,.$
} $V$. 
Under these conditions, it can be shown that the coefficients of the Chebyshev interpolant of $f(x)$ satisfy
\begin{equation}
    |\hat{a}_k| \leq \frac{2V}{\pi (k - m)^{m+1}},
\end{equation}
for all $k > m$.
Moreover, there is a well-behaved bound on the interpolation error associated to $p_n(x)$. 
Using this, under the previous conditions, it can be shown that $p_n(x)$ satisfies
\begin{equation}
    \|f(x) - p_n(x)\|_\infty \leq \frac{4V}{\pi m (n-m)^m}, \qquad n  > m,
\end{equation}
where $\|g(x)\|_\infty$ is the $\sup$-norm of $g(x)$ on the range $[-1,1]$, i.e.
\begin{equation}
    \|g(x)\|_\infty = \sup\{|g(x)| \,\, : \,\, x \in [-1, 1]\}.
\end{equation}
That is, the largest error across the interpolation range falls algebraically with increasing $n$.

The algebraic rate of convergence for differentiable functions is greatly improved
when one considers analytic functions instead. In this case, the relevant result states
that if $f(x)$ is analytic on $[-1,1]$ and admits an analytic continuation to the 
open Bernstein ellipse $E_\rho$, and if $|f(z)|< M$ for some $M>0$ and all $z \in E_\rho$, then its 
Chebyshev coefficients satisfy
\begin{equation}
    |\hat{a}_0| \leq M, \qquad \quad |\hat{a}_k| \leq 2M\rho^{-k},
    \label{eq:coefficientsGeometric}
\end{equation}
for all $k \geq 1$,
while the interpolation error in the $\sup$-norm satisfies
\begin{equation}
     \| f(x) - p_n(x) \|_{\infty}\leq \frac{4M\rho^{-n}}{\rho-1}
     \label{eq:AnalyticChebyshevErrorBound}
\end{equation}
for all $n  \geq 0$.
That is, Chebyshev interpolation achieves geometric convergence for analytic 
functions across the interpolation range\footnote{For entire functions, one finds super-geometric convergence since 
the Bernstein parameter $\rho$ can be taken arbitrarily large, leading to decay 
faster than any fixed geometric sequence.}. Eq.~\eqref{eq:AnalyticChebyshevErrorBound} 
is the main motivation for exploring the use of Chebyshev polynomials to approximate 
Feynman integrals: if we restrict to a segment of phase-space inside of the well
understood regions of analyticity, 
then we find that we can obtain a useful approximation with comparatively low approximation 
order.

\begin{figure}[htb!]
    \centering
    \includegraphics[scale=0.8]{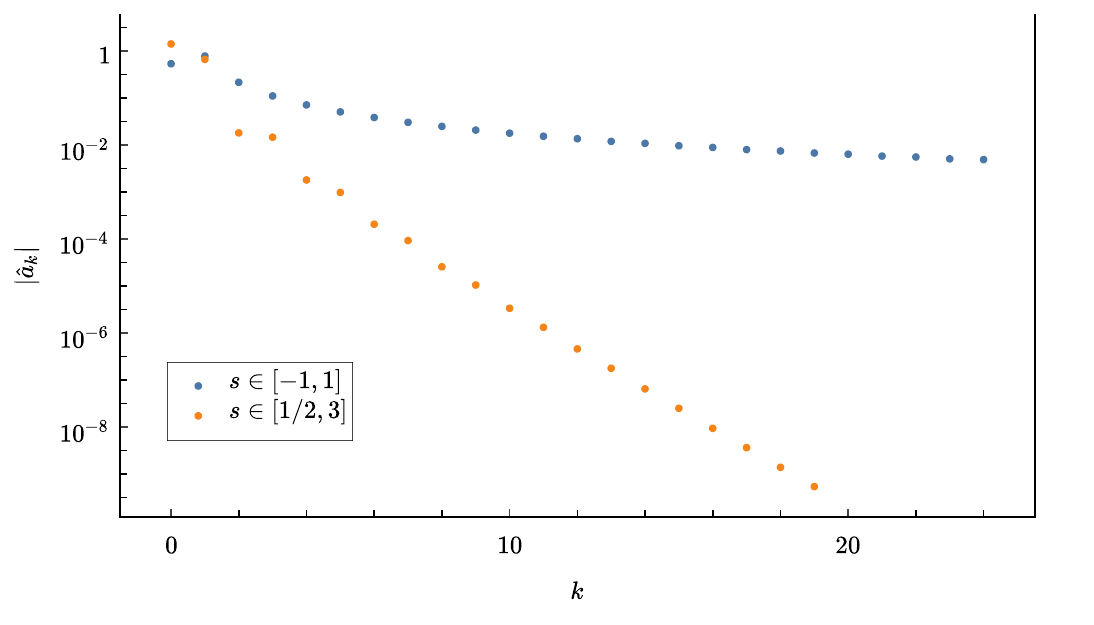}
    \caption{$|\hat{a}_k|$ versus $k$ for $I_{\text{bub}}(s, m^2 = 1)$ over the intervals $s\in[-1,1]$ (blue) and $s\in[1/2,3]$ (orange) on a log scale. The flattening behavior of the first interval can be attributed to algebraic convergence, while the straight line behavior of the second interval can be attributed to geometric convergence.}
    \label{fig:Chebyshev_Coefficients_Example}
\end{figure}

To illustrate the convergence properties of Chebyshev interpolants, let us return to the example of the bubble integral in \cref{eq:Bubble_Integral}. We interpolate $I_{\text{bub}}(s, m^2 = 1)$ over two intervals $s \in [-1,1]$ and $s \in [1/2,3]$ as shown in \cref{fig:Chebyshev_Coefficients_Example}.
Since $I_{\text{bub}}(s,1)$ is analytic in a neighbourhood of $[1/2, 3]$, the corresponding Chebyshev expansions exhibit geometric convergence. This is reflected in the asymptotically linear decay of the expansion coefficients on a logarithmic scale. 
In contrast, $I_{\text{bub}}(s,1)$ is continuous but not differentiable at $s=0$. Hence, the interpolant on $[-1,1]$ exhibits only algebraic convergence. This is evidenced by a departure from linear behaviour, with coefficients bending upward, away from a straight line, with a slope that tends to zero from below.

\subsection{Working with Chebyshev Series}
\label{sec:ChebyshevOperations}

Having motivated that Chebyshev series constructed via interpolation provide a valuable method of approximation for 
Feynman integrals, let us now discuss the necessary tools for working with them to achieve our goals of approximating the integrals in \cref{eq:PathIntegrationForm}.
In practice, this set is quite limited which adds to the attractiveness of the approach. 
We must know how to construct a Chebyshev series via interpolation, compute the anti-derivative and evaluate 
a Chebyshev series to match the boundary condition in \cref{eq:PathIntegrationForm}.

\paragraph{Constructing a Chebyshev Interpolant}
Let us consider the problem of computing an order $n$ Chebyshev interpolant $p_n(x)$ for a given function $f(x)$. That is, we wish to construct the set of coefficients $\hat{a}_k$ of \cref{eq:ChebyshevInterpolantForm}.
Mathematically, this can be achieved by leveraging the discrete orthogonality condition satisfied by the 
Chebyshev polynomials \cite{Chebyshev_Review},
\begin{equation}
    \sum_{k=0}^{n}{}^{\prime\prime} T_i(x_k) T_j(x_k) =  \begin{cases}
        0, \qquad i\neq j \quad \text{and} \quad i, \, j \leq n\\
        \frac{n}{2}, \qquad i = j \quad \text{and} \quad 0 < i <n \\
        n, \qquad i=j \quad \text{and either} \quad i= 0 \quad \text{or} \quad i=n,
        \end{cases}
        \label{eq:Discrete_Orthogonality}
\end{equation}
where the $x_k$ are the Chebyshev-Lobatto nodes and the double
prime notation on the summation indicates that both the first and the last terms in the sum are
to be halved.
Given \cref{eq:Discrete_Orthogonality}, it follows that
\begin{equation}
         \hat{a}_j = \frac{2}{w_j n}  \sum_{k=0}^{n}{}^{\prime\prime} T_j(x_k)f(x_k),
     \label{eq:Coefficient_Expression_1}
\end{equation}
where we make use of $w_j$ defined as
\begin{equation}
    w_0 = w_n = 2, \quad \text{and} \quad w_j = 1 \quad \text{otherwise}.
\end{equation}

While one can take \cref{eq:Coefficient_Expression_1} as a direct prescription for computing the interpolant coefficients, this can be further developed by considering \cref{eq:Coefficient_Expression_1} more closely. First, we note that \cref{eq:Coefficient_Expression_1} can be considered as a linear transformation acting on the vector of evaluations of $f(x)$ at the Chebyshev-Lobatto nodes. Specifically,
\begin{equation}
    \hat{a}_j  = \sum_{k=0}^{n} \mathcal{I}_{jk} f_k
\end{equation}
where we have introduced the shorthand notation $f(x_k)=f_k$ and $\mathcal{I}_{jk}$ is an $(n+1)\times(n+1)$ interpolation matrix. The entries of $\mathcal{I}_{jk}$ are given by
\begin{equation}
    \mathcal{I}_{jk} = \frac{2}{w_j w_k n}\cos\left(\frac{jk\pi}{n}\right),
\end{equation}
where we have made use of \cref{eq:Chebyshev_Trignometric_Argument,eq:Chebyshev_Extrema} to re-express the 
evaluation of the Chebyshev polynomials directly in terms of the cosine function.
Importantly, up to normalization factors, we see that the matrix $\mathcal{I}_{jk}$ corresponds 
to the action of a so-called ``discrete cosine transform'' (DCT) of type 1. That is, the Chebyshev coefficients 
may be written as:
\begin{equation}
         \hat{a}_j = \frac{2}{w_j n}  \left[\frac{f_0}{2} + \sum_{k=1}^{n-1}\cos\left(\frac{jk\pi}{n}\right)f_k  + (-1)^j\frac{f_n}{2}\right].
    \label{eq:Coefficient_Expression_DCT}
\end{equation}
In practice, the DCT can be implemented efficiently using so-called ``fast cosine transform'' approaches. One way to achieve this 
is to rephrase the DCT in terms of a discrete Fourier transform.
To that end, we begin by defining the extended vector $F_k$ of length $2n$ in the following manner:
\begin{equation}
        F_k = \begin{cases}
        \Tilde{f}_k, \quad 0 \leq k \leq n\\
        \Tilde{f}_{2n-k}, \quad n<k<2n
    \end{cases}
\end{equation}
where
\begin{equation}
    \Tilde{f}_k = \frac{1}{w_k} f_k.
\end{equation}
The extended sequence is, by construction, even-symmetric, ensuring that only cosine modes contribute in the Fourier expansion. Consequently, we may calculate the Chebyshev interpolant coefficients in terms of a discrete Fourier transform as
\begin{equation}
         \hat{a}_j = \frac{1}{w_j n}  \sum_{k=0}^{2n-1} F_k e^{-2\pi i k j/2n},
    \label{eq:Coefficient_Expression_FFT}
\end{equation}
which can naturally be evaluated efficiently using a fast Fourier transform (FFT) algorithm.

To close, let us consider the efficiency of each approach. 
The interpolation-matrix approach requires $\mathcal{O}(n^2)$ operations. 
In contrast, the DCT and FFT-based approaches require only $\mathcal{O}(n\log(n))$ operations. 
While for sufficiently large $n$ the DCT/FFT-based approach is generally more efficient, for the 
moderate values of $n$ encountered in Chebyshev approximation, the performance differences 
between these approaches are more strongly affected by the specifics of the implementation.

\paragraph{Evaluating a Chebyshev Series}
\begin{algorithm}[t]
\caption{Clenshaw's Recurrence}\label{alg:clenshaw}
\KwData{coefficients $a_0,\dots,a_n$, evaluation point $x$}
\KwResult{$y = \sum_{k=0}^n a_k T_k(x)$}

$b_{n+1} \gets 0$\;
$b_{n+2} \gets 0$\;

\For{$k \gets n$ \KwTo $0$ step $-1$}{
    $b_k \gets a_k + 2x\,b_{k+1} - b_{k+2}$\;
}

$y \gets b_0 - x b_1$\;

\Return $y$\;

\end{algorithm}

Once the Chebyshev series has been obtained, the next step is its evaluation at a given point $x\in[-1,1]$. 
Direct evaluation of the series can become numerically unstable for large values of $n$ due to round-off 
and cancellation errors. Clenshaw's recurrence algorithm \cite{Clenshaw} is a computationally stable 
and efficient algorithm for evaluation, requiring $\mathcal{O}(n)$ operations.
The algorithm can be applied to any family of functions satisfying a three-term recurrence relation.
In the case of Chebyshev polynomials, they naturally satisfy a recursion relation given by 
    \begin{equation}
        T_{n+1}(x) = 2x T_{n}(x) - T_{n-1}(x),
    \end{equation}
where the base case of the recursion is given by
\begin{equation}
    T_0(x) = 1, \quad \quad T_1(x) = x,
\end{equation}
which follows directly from \cref{eq:Chebyshev_Trignometric_Argument}. One can thus make use of 
Clenshaw's algorithm which we describe in \cref{alg:clenshaw}.
For a comprehensive discussion of Clenshaw's algorithm, we refer the reader to 
standard texts in numerical analysis, such as refs.~\cite{Numerical_Analysis_1,Numerical_Analysis_2}.

\paragraph{Integrating a Chebyshev Series}

In the context of solving differential equations for Feynman integrals, our approach will be to 
construct a Chebyshev interpolant for the integrand in \cref{eq:PathIntegrationForm}, and to then 
integrate up this interpolant to find a Chebyshev approximation for the Feynman 
integral itself. As such, we need to be able to compute the Chebyshev series of 
the integral of some other Chebyshev series.
Naturally, as Chebyshev polynomials form a basis of the vector space of all polynomials,
the action of operators such as differentiation and integration of a Chebyshev
polynomial may be represented as a linear combination of Chebyshev polynomials.
Due to the connection of Chebyshev polynomials to trigonometric functions, i.e. \cref{eq:Chebyshev_Trignometric_Argument}, 
it is not hard to see that integration of an arbitrary Chebyshev polynomial can be achieved 
in closed form. Specifically, it can be shown that
\begin{equation}
    \int T_n(x) dx = \begin{cases}
        \frac{1}{2}\left[\frac{T_{n+1}(x)}{n+1}-\frac{T_{|n-1|}(x)}{n-1}\right] + C, \quad n \neq 1, \\
        \frac{1}{4}T_2(x) + C, \qquad \qquad \qquad \,\,\, \quad n=1,
    \end{cases}
    \label{eq:Chebysev_Integration}
\end{equation}
for some constant of integration $C$.
Similar results for differentiation can be found in ref.~\cite{Chebyshev_Review}. 
Thus we see that the indefinite integral of a Chebyshev series is itself a Chebyshev series of degree one 
higher than the original series. 
Using \cref{eq:Chebysev_Integration}, the coefficients of the antiderivative can be related to 
those of the original series as:
\begin{equation}
    \int  \sum_{k=0}^n a_kT_k(x) \, \mathrm{d} x=  C +  \sum_{k=1}^{n+1} A_kT_k(x), 
    \qquad 
    A_k = 
    \begin{cases}
        a_0 -\frac{a_2}{2}, \quad \quad k=1 \\
         \frac{a_{k-1}-a_{k+1}}{2k}, \quad k>1,
    \end{cases}
\end{equation}
where we implicitly take $a_{n+1} = a_{n+2}=0$. As such, it is clear that integration of a (finite) Chebyshev series is a highly
efficient operation that takes only $\mathcal{O}(n)$ operations.
Although not directly relevant for our goal, we remark that an analogous result can 
be obtained for the derivative of a Chebyshev series \cite{chebfun1}, which can be useful for cross-check purposes.

\subsection{Numerical Convergence of Chebyshev Series}
\label{sec:PracticalConvergence}

\begin{figure}[tb]
    \centering
    \includegraphics[width=1\linewidth]{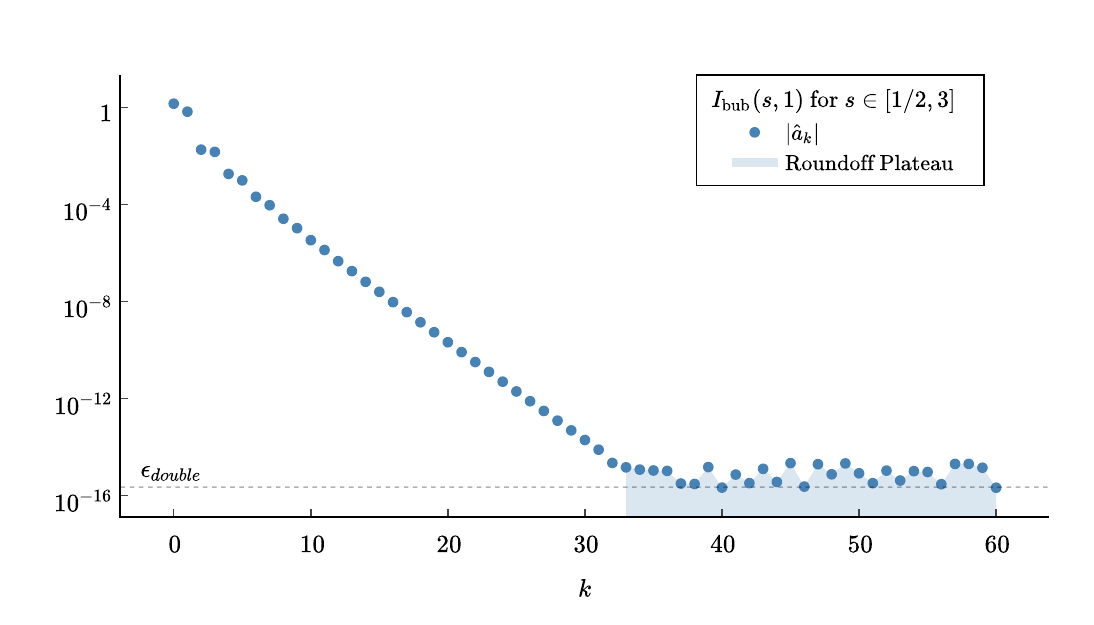}
    \caption{$|\hat{a}_k|$ versus $k$ for $I_{\text{bub}}(s, m^2 = 1)$ over the interval $s\in[1/2,3]$. The blue shaded region corresponds to the roundoff plateau for which roundoff errors associated with machine precision dominate the coefficient calculation. In this region, the coefficients oscillate around the roundoff error scale.}
    \label{fig:roundoffPlateau}
\end{figure}

In order to practically build a Chebyshev interpolant for a univariate function,
an important decision to make is to understand the order $n$ of the Chebyshev approximation or, 
correspondingly, the number of nodes used for the collocation. As, in this work, we wish to apply 
Chebyshev collocation methods to analytic functions, it is natural to employ the error bound \cref{eq:coefficientsGeometric}
to guide this decision. In principle, this requires knowledge 
of the Bernstein radius of analyticity, $\rho$ as well as the scale-setting constant $M$. While this 
could systematically be extracted from analytic knowledge of the function that we are interpolating, 
this requires case-by-case analysis. For this reason, we follow a numerical approach to understanding 
the convergence of a Chebyshev interpolant~\cite{Aurentz2017}, as used in the 
\texttt{chebfun} project~\cite{chebfun1,Chebfun}. 

Importantly, while the coefficients of the Chebyshev series of an analytic
function exhibit geometric decay,
in practical applications of Chebyshev series, one is confronted by numerical challenges in determining these coefficients. Indeed, as the coefficient magnitudes decrease, they inevitably reach the scale of the precision of the input data, beyond which the coefficients can no longer be computed accurately. At this stage, roundoff errors become the dominant source of error. 
This regime is known as the roundoff plateau. Within this region, the coefficients no longer decay exponentially but rather fluctuate randomly around the roundoff error scale.
As the coefficients in the roundoff plateau are dominated by numerical noise, their computation offers little to no benefit. 
The task is thus to identify, in an automatic and function-agnostic manner, the point beyond which the coefficients become corrupted by roundoff errors and to truncate the expansion accordingly.

Our approach is motivated by the observation that, when the absolute value of the expansion coefficients is plotted on a logarithmic scale, the roundoff plateau manifests itself as a region of relative flatness, with the coefficients oscillating about a median value (see \cref{fig:roundoffPlateau}). If a straight line were to be fitted to the coefficients within this region, the resulting slope would be close to zero. 
By contrast, coefficients that remain relatively unaffected by roundoff errors are expected to display a strong negative slope corresponding to their exponential decay. Therefore, the task of identifying the roundoff plateau reduces to locating regions for which the local slope becomes sufficiently small.
In what follows, we formalise this intuitive picture and provide a robust numerical implementation.

\begin{figure}
    \centering
    \includegraphics[width=1\linewidth]{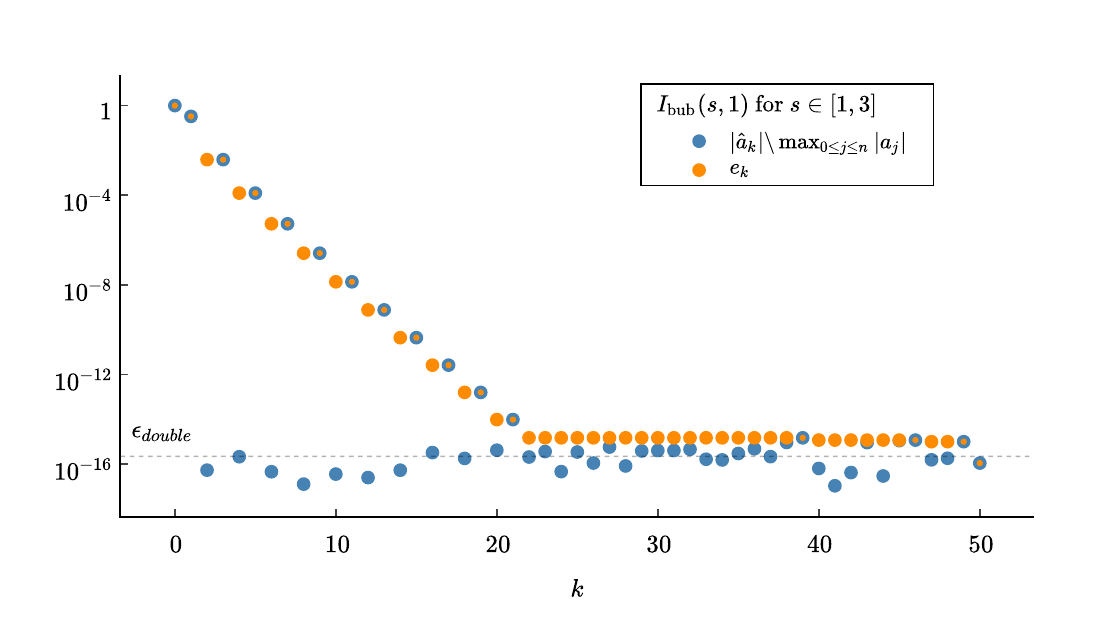}
    \caption{A depiction of the envelope of the sequence of (normalised) Chebyshev interpolant coefficients of $I_{\text{bub}}(s, m^2 = 1)$ on the interval $s\in[1,3]$ for $n=50$, alongside the normalised Chebyshev series. Bicoloured points denote points where the envelope and normalised coefficients are equal.
    The envelope acts as a filter, smoothing the oscillation of the expansion coefficients.}
    \label{fig:envelope}
\end{figure}

We begin by introducing the notion of an envelope of a sequence $a_k$, defined as~\cite{Aurentz2017}
\begin{equation}
    e_j = \max_{j \leq k \leq n} |a_k|.
\end{equation}
Provided that $e_0 \neq 0$, the sequence is normalised by $e_0$, resulting in a non-negative, monotonically non-increasing sequence whose first element equals $1$. The introduction of the envelope is motivated by two considerations. 
First, coefficients in a spectral expansion may exhibit oscillatory behaviour. For example, due to symmetry, all odd Chebyshev coefficients in the expansion of an even function vanish identically. 
The envelope filters these oscillations, resulting in a smoother sequence while preserving the relevant underlying structure.
Second, the envelope normalisation renders the algorithm scale-invariant, thereby mitigating the need to operate with coefficients whose magnitudes may potentially span several orders of magnitude.

\begin{figure}
    \centering
    \includegraphics[width=1\linewidth]{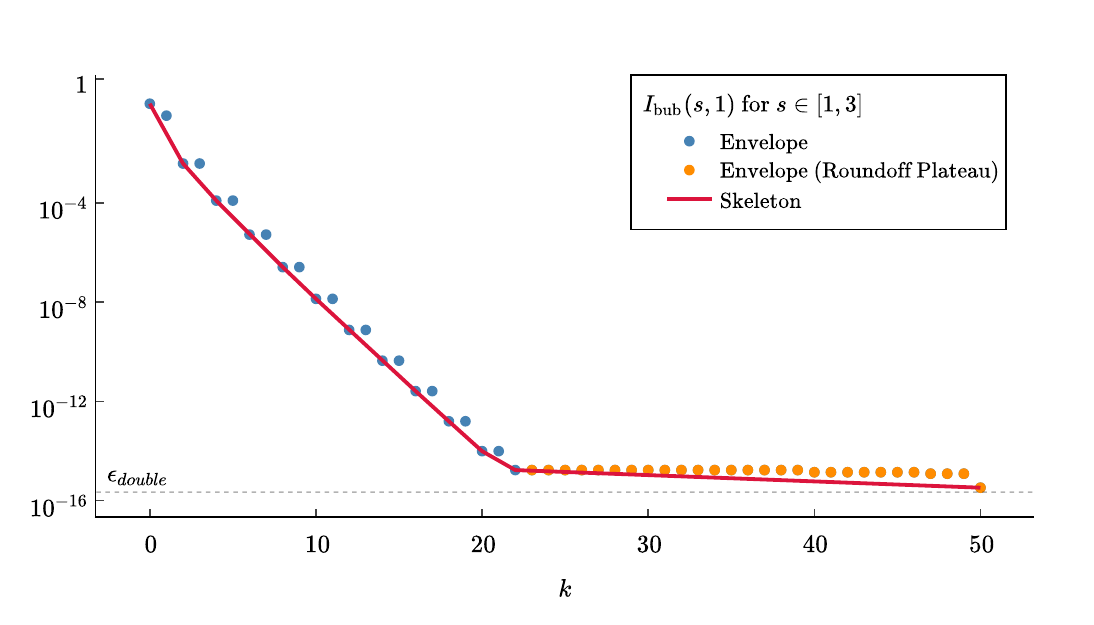}
    \caption{A depiction of the envelope of the sequence of Chebyshev interpolant coefficients of $I_{\text{bub}}$ for $n=50$, alongside its logarithmic skeleton. The logarithmic skeleton clearly provides a simple, nearly linear approximation to the envelope. The roundoff plateau (in orange) is clearly identified as a long flat stretch. Due to the simplicity of the input function, the roundoff plateau is controlled by the precision of machine floating point numbers.}
    \label{fig:skeleton}
\end{figure}

In our applications, the important benefit of working with the envelope is that it allows for a practical 
way to identify the convergence rate of a Chebyshev series.
When considering examples such as that in \cref{fig:envelope}, it is clear that the
(logarithm) of the sequence can be approximated
by a collection of straight lines.
If one is able to robustly identify such a continuous model, this allows one to
construct a simple notion of a derivative at any (even non-integer)
point.\footnote{We note that this discussion represents a
departure from the approach of ref.~\cite{Aurentz2017}, principally motivated by the physicist's desire to work with continuous objects.}
To this end, we draw inspiration from recent applications of tropical geometry
and Newton polygons/polytopes in the study of scattering amplitudes (see
e.g.~\cite{Salvatori:2024nva, Giroux:2026tgd, Borinsky:2020rqs,
  Borinsky:2023jdv, Heinrich:2021dbf, Pak:2010pt, Arkani-Hamed:2022cqe}).
Specifically, to attach a continuous linear model to the
logarithm of our envelope, we take its lower convex hull. As we work in two
dimensions, this can be neatly expressed as
\begin{equation}
    s(x) = \min_{\substack{i \ne j \\ i\le x\le j}} \left( \frac{j-x}{j-i} \log_{10}(e_i) + \frac{x-i}{j-i} \log_{10}(e_j) \right),
\end{equation}
where the minimum is taken over $i$ and $j$.
We refer to $s(x)$ as a logarithmic ``skeleton'' of the envelope, that coarsely captures the structure of the non-increasing sequence.
The skeleton is a piecewise linear function in log-space, whose graph is the lower
facets of a convex polygon. The vertices of these facets can be efficiently
computed from the envelope itself in $\mathcal{O}(n)$, for example, using the
monotone chain algorithm.
Importantly, as the skeleton is convex, its derivative is also a non-increasing
function. To exemplify this, in \cref{fig:skeleton}, we depict the envelope and logarithmic skeleton of the
envelope for a Chebyshev approximation of the $I_{\text{bub}}(s, 1)$ on $s \in [1, 3]$. One can clearly see in this case that it is easy to identify from the
skeleton where one reaches the roundoff plateau.

To practically define the location of the roundoff plateau we use a cutoff strategy, defining it as an $x_{\text{plateau}} \in [0, \tilde{n}]$ such that
\begin{equation}
    s'(x_{\text{plateau}}) < \nabla_{\text{cutoff}},
\end{equation}
where $\tilde{n} < n$. We make use of $\tilde{n}$ to ensure that the plateau is stable, and not 
only reached in the final coefficients, taking a value of $\tilde{n} = 0.85 n$.
In practice, we choose a gradient cutoff of $\nabla_{\text{cutoff}} = 0.02$.
Beyond this, we only accept the plateau if the sum of the last five elements of the envelope sequence is 
less than $10^{-6}$ in order to explicitly ensure that the envelope has
reached the asymptotic regime.
If there is no $x_{\text{plateau}}$ which satisfies these conditions, we conclude that the
Chebyshev series has not yet converged. We will later use this ``plateau test''
to test for convergence in our adaptive algorithm. We note that ref.~\cite{Aurentz2017} uses a more sophisticated
plateau detection criterion. While this would be interesting to investigate, we find that the simple
cutoff approach is sufficient for phenomenological application and leave such investigations for further work.

Beyond using the plateau location to determine if a Chebyshev series has numerically converged, we can also use
it to estimate the convergence rate. As discussed in \cref{sec:Chebysev_Series}, for an analytic function, the gradient of the
coefficient curve on a log-scale plot approximates $\log_{10}(\rho)$. Let us define $n_{\text{plateau}} = \text{floor}(x_{\text{plateau}})$
as the location in the sequence just before the plateau.
We define an approximation to $\rho$ as
\begin{equation}
    \log_{10}(\rho_{\text{approx}}) = -\log_{10}(e_{n_{\text{plateau}}})/n_{\text{plateau}}.
\end{equation}
This quantity, on the log-scale plot of the envelope gives some average derivative of the curve, thereby giving a
coarse measure of the rate of convergence of the sequence. In our adaptive approach, $\rho_{\text{approx}}$ will
be useful to detect if a Chebyshev series on a particular range is converging at a sufficient rate and to act accordingly.

\subsection{An Adaptive Chebyshev Approach for Feynman Integrals}
\label{sec:AdaptiveAlgorithmDescription}

Having introduced a sufficient collection of Chebyshev technology let us now discuss how we 
can use this to construct a robust algorithm to compute the integrals in \cref{eq:PathIntegrationForm}.
Our approach is to construct a Chebyshev
approximant for $\tilde{I}^{(n)}_i(\vec{s}[t])$ on the segment $t = [0, 1]$, in an
iterative fashion. We assume the existence of a numerical black-box procedure
for the evaluation of $\tilde{I}^{(n-1)}_i(\vec{s}[t])$. For $n=1$, this is given by the
weight 0 solutions to the integrals, which are rational numbers and easily
computed analytically. For $n > 1$, we make use of the output of our procedure at a previous iteration.
At fixed $n$, we construct a Chebyshev interpolant of the
integrand in \cref{eq:PathIntegrationForm}. 
Given the discussion in \cref{sec:ChebyshevOperations}, we then perform the integration of the 
interpolant analytically, arriving at a Chebyshev approximation of 
the anti-derivative in \cref{eq:PathIntegrationForm}. 
Finally, we fix the constant of integration to match the boundary value $\tilde{I}^{(n)}_i(\vec{s}_0)$. 
We use Clenshaw's algorithm to evaluate the Chebyshev approximant at
$\vec{s}_0$, finally arriving at a Chebyshev approximation for $\tilde{I}_i^{(n)}(\vec{s}[t])$ which can efficiently be evaluated anywhere along the path using Clenshaw's algorithm.

The main ingredient in our procedure is thus to construct a Chebyshev interpolant of the integrand in \cref{eq:PathIntegrationForm}. 
Our approach is ``adaptive'': we use the technology of \cref{sec:PracticalConvergence} to robustly stop sampling in the presence of roundoff error, 
and increase sampling density in the presence of nearby non-analyticities.
For generality, we denote the function under approximation as $f(t)$ and the Chebyshev approximation range as $[t_0, t_1]$.
Similar to commonly used series expansion methods (e.g.~\cite{Moriello:2019yhu}), we achieve this by taking the original range and 
splitting it up into ``segments''.
We adaptively choose a segmentation based on the observed convergence properties of the Chebyshev approximation, leading to more dense sampling in regions of slower convergence.

A first consideration that must be discussed is the choice of map from a segment $[T_0, T_1]$ 
in $t$ space to the range $[-1, 1]$ in Chebyshev space. 
%Here we discuss generalities, and return to the specific choice of parameters in \cref{sec:FivePointPhaseSpace}.
%
Let us denote the Chebyshev variable by $x$. In practice, a map from $t$-space to $x$-space is simply an expression for 
$x$ as a function of $t$. Analogously, this map can be inverted and we can consider $t$ as a function of $x$.
The simplest choice of map to Chebyshev space is an affine map. The affine map from $[-1, 1]$ to $[T_0, T_1]$, 
alongside its inverse are given by
\begin{align}
    t(x) &= T_1 - (T_1 - T_0) \frac{1 - x}{2},
    \label{eq:AffineFromChebyMap}
    \\
    x(t) &= 1 - 2 \frac{T_1 - t}{T_1 - T_0},
    \label{eq:AffineToChebyMap}
\end{align}
which is uniquely fixed by the fact that the map is linear, together with the requirement that 
$x(T_0) = -1$ and $x(T_1) = 1$.

While the affine map is simple, it can be useful to consider other choices of map to improve the convergence of 
Chebyshev approximation. 
Recalling the discussion in \cref{sec:Chebysev_Series}, we note that the radius of the Bernstein ellipse can 
be limited by the presence of an algebraic branch point of $f(t)$ close to the segment $[T_0, T_1]$. 
Indeed, in practical applications to Feynman integrals, it frequently occurs that there is a square-root 
branch point at some $t_2 \not \in [t_0, t_1]$
that can be associated with the vanishing of a Gram determinant. 
A natural solution to this is to choose a map to Chebyshev space such that, as a function of $x$, the argument of the square-root is locally a perfect square.
Colloquially, we will say that this map ``resolves'' the square root branch point.
A map from the segment $[T_0, T_1]$ to $[-1, 1]$ that resolves a square root branch point at $t_2 \not\in [t_0, t_1]$ is 
\begin{align}
t(x) &=  t_2 - (t_2 - T_0)\left(1 - \frac{1 - u(T_1)}{2}(x + 1)\right)^2, 
\label{eq:AlgebraicFromChebyMap}
\\
x(t) &= 2\frac{1 - u(t)}{1 - u(T_1)} -1,
\label{eq:AlgebraicToChebyMap}
\end{align}
where
\begin{equation}
u(t) = \sqrt{\frac{T_2 - t}{T_2 - T_0}}.
\end{equation}
When using such a non-linear map to approximate $f(t)$ in a context where one desires to compute the anti-derivative, 
one is faced with the question of what to do with the Jacobian of the map.
As it must eventually be included by the time of integration, when using a non-linear map, we choose to construct an approximation 
for $f(t)$ including the Jacobian factor. 
This choice has an important side effect. Specifically, if the square root singularity at $t = t_2$ is integrable, this behaviour 
is cancelled by the inclusion of the Jacobian factor.
In practice, this allows our technology to interpolate functions with integrable singularities.

\SetKwBlock{RefineBlock}{\textbf{Refine Segment:}}{}

\begin{algorithm}[t]

\DontPrintSemicolon
\KwIn{Function $f(t)$, range $[t_0, t_1]$}
\BlankLine
Set initial segment to full range, i.e. $[T_0, T_1] \leftarrow [t_0, t_1]$\;
\BlankLine
\nl\label{step:start}\textbf{Segment Initialise:} Sample the segment on a mapped $n = 8$ Chebyshev-Lobatto grid and build the interpolant\;
\BlankLine
\nl\label{step:refinement}\RefineBlock{
    \lIf{$\rho_{\mathrm{approx}} < \rho_0$}{\textbf{go to} \ref{step:bisect}}
    \lIf{Converged by plateau test}{\textbf{go to} \ref{step:nextSegment}}
    \lIf{$n < 128$}{Set $n \leftarrow 2 n$, sample to build the interpolant, and \textbf{go to} \ref{step:refinement}}
}
\BlankLine
\nl\label{step:bisect}\textbf{Bisect Segment:} Set $T_1 \leftarrow T_0 + (T_1 - T_0)/2$ and \textbf{go to} \ref{step:start}
\BlankLine
\nl\label{step:nextSegment}\textbf{Next Segment:} \If{$T_1 \neq t_1$}{
    Move to the remaining part of the range, setting $[T_0, T_1] \leftarrow [T_1, t_1]$\;
    \textbf{go to} \ref{step:start}\;
}

\caption{Adaptive Chebyshev Interpolation}
\label{alg:Adaptive}
\end{algorithm}

Let us now discuss our algorithm for constructing a Chebyshev approximation of 
a function $f(t)$ on a range $[t_0, t_1]$. We present the algorithm in
\cref{alg:Adaptive}. 
Starting with a tentative initial segment $[T_0, T_1] = [t_0, t_1]$, we sample on a small Chebyshev-Lobatto grid. 
Computing the associated Chebyshev coefficients, we then use these to decide how to proceed. 
We first check if the Chebyshev approximation seems to be converging. To this end, we compute $\rho_{\text{approx}}$ and 
check if it is above some target value $\rho_0$. If the convergence is not strong enough, we decide to bisect the segment. 
If the convergence behaviour is satisfactory, we next check to see if the approximation has already converged by the plateau test. 
If the approximation has not converged, we double the Chebyshev order and resample. Importantly, as the Chebyshev-Lobatto 
grid for an interpolant of degree $2n$ contains the grid for degree $n$, one can recycle the known evaluations.
We repeat this until either the approximation converges or we hit a cutoff of $128$ samples. At this stage, 
we move on to process the remaining segment, or terminate.

The output of this approach is thus a collection of Chebyshev approximations of
$\tilde{I}^{(n)}_i(\vec{s}[t])$, one for each segment, that can easily be evaluated at any point along the path. 
It thus allows us both to evaluate the integrals at the point of interest, $\vec{s}_1$ as well as provides
input to the calculational procedure at weight $n+1$.

We close with some technical comments on our implementation.
Firstly, we note that \cref{alg:Adaptive} requires the specification of
a ``target'' convergence rate $\rho_0$. In our implementation, we choose
$\log_{10}(\rho_0) = 0.2$. This is motivated by the ad-hoc desire to achieve an
increase of precision of 1 digit for every 5 coefficients, therefore potentially
reaching machine precision in around $80$ evaluations. The sample cutoff of $128$ 
in \cref{alg:Adaptive} is motivated as the smallest power of 2 beyond this expectation. 
In future work it would be interesting to understand the impact of the choice of 
$\rho_0$ on performance. Nevertheless, in practical applications we find this choice acceptable.
Secondly, in practical implementation of our approach, sufficiently degenerate phase
space points can be challenging. 
To ensure that the algorithm does not continue
indefinitely due to such precision issues, we limit the number of segments to 25
and a minimal segment length of $10^{-7}(t_1 - t_0)$, and require that the final
such segment always uses 128 samples.
As a final comment, we note that an affine-mapped Chebyshev-Lobatto grid of order $n$ covering $[T_0, T_1]$ and $[T_0, T_0 + (T_1 - T_0)/2]$ only share a small number of nodes.
Thus, when the path is bisected, previous evaluations cannot be efficiently reused.
This motivates our early and repeated detection of convergence behaviour.

\section{Application to Pentagon Functions}
\label{sec:pentagonFunctions}

A major aim of this work is to develop a general tool for solving canonical 
differential equations such as \cref{eq:CanonicalDE} that is not tailored to a specific problem and is suitable for phenomenological applications. 
In this section, we stress test our approach by 
applying it to recompute various collections of two-loop five-point Feynman 
integrals. Specifically, we consider the numerical evaluation of pentagon 
functions for full-colour five-point massless processes and leading-colour five-point 
one-mass processes, which were first presented in a form suitable for phenomenological 
application in ref.~\cite{Chicherin:2020oor} and ref.~\cite{Chicherin:2021dyp} respectively.
In this section, we describe details of the application of our adaptive Chebyshev 
approximation approach, discuss its performance 
over phase-space, and provide a proof of concept \texttt{Mathematica} implementation in ancillary files \cite{ancfiles}.

\subsection{Physical Phase Space for Five-Point Processes}
\label{sec:FivePointPhaseSpace}
In full generality, five-point Feynman integrals
are functions of five momenta $p_i$, which satisfy total momentum conservation,
i.e. $\sum_{i=1}^5 p_i^\mu = 0$, where we make use of the all incoming
convention.
The specifics of the on-shell conditions depends on the particular process under
consideration, which we return to later.
We are interested in evaluating these collections of Feynman integrals at a point
in the physical scattering channel
\begin{equation}
  p_1, p_2 \rightarrow p_3, p_4, p_5.
  \label{eq:PhysicalChannel}
\end{equation}
We will achieve this by solving their differential equations by evolution
over a path that lives entirely within physical phase space.

While physical phase space is most naturally described in momentum space,
where scattering momenta that conserve total momentum are
on-shell and real-valued and have the appropriate signs of their energy
components, it is frequently the case that the differential equations are
solved in the space of Mandelstam variables, $s_{ij} = (p_i + p_j)^2$ (and $p_i^2$ 
if external legs are massive).
In terms of Mandelstam variables, the momentum-conservation and on-shellness
constraints are easily solved, while the energy positivity constraints give rise
to linear inequalities that the Mandelstam variables must satisfy.
In contrast, the reality condition for the momenta translates into a non-trivial
constraint on the Gram determinant associated to the five momenta,
\begin{equation}
  G_5 = \det(\{p_i \cdot p_j\}_{i,j \in \{1, \ldots, 4\}}).
\end{equation}
In physical phase space, it is well known that the Gram determinant is negative~\cite{Byers:1964ryc}.
That is,
\begin{equation}
  G_5 < 0,
  \label{eq:GramNegativity}
\end{equation}
which follows from the physical metric signature.

In order to practically solve the differential equation, given an initial point $\vec{s}_0$ and 
a final point $\vec{s}_1$, we must construct a path that lives entirely within physical phase-space.
For simplicity, it is easiest to work with straight-line paths in Mandelstam space. That is, we consider a path of the form
\begin{equation}
    \vec{s}[t] = \vec{s}_a (1-t) + \vec{s}_b t.
    \label{eq:StraightLinePath}
\end{equation}
However, it is well understood that the Gram negativity constraint, \cref{eq:GramNegativity}, cuts out a 
non-convex, often not star-shaped region of Mandelstam space~\cite{Chicherin:2020oor, Chicherin:2021dyp}.
In order to overcome this problem, we follow the approach of~\cite{Chicherin:2021dyp}. Specifically, we construct a
path from $\vec{s}_0$ to $\vec{s}_1$ as a concatenation of straight-line paths, 
each of which lives entirely within the region of phase space under consideration. To do this, we first check if the straight line between 
$\vec{s}_0$ and $\vec{s}_1$ lives within physical phase space. If not, we repeatedly generate an intermediate 
physical point $\vec{s}^*$ until the straight lines $\vec{s}_0 \rightarrow \vec{s}^*$ and 
$\vec{s}^* \rightarrow \vec{s}_1$ live within physical phase space. As expected from previous computations~\cite{Chicherin:2020oor, Chicherin:2021dyp,Abreu:2023rco}, we can always construct such a two-segment path, i.e.~we can always find such a point $\vec{s}^*$.

The Gram determinant $G_5$ also plays an important role in the analytic structure of five-point Feynman integrals. 
Specifically, the collection of Feynman integrals
that we study in this work have an algebraic branch point at $G_5 = 0$. As this is
the boundary of phase space, and we choose to solve the differential equation
along paths entirely within physical phase space, we will never cross this
branch point.
Nevertheless, in physical phase-space samples, points where $G_5$ is
small can frequently be found, which has an important impact on the convergence 
properties of our approach.
We use this observation to guide the choice of mapping to Chebyshev space that was 
discussed in \cref{sec:AdaptiveAlgorithmDescription}.
Specifically, we use the square-root branch-point resolving map of 
\cref{eq:AlgebraicFromChebyMap,eq:AlgebraicToChebyMap}, choosing the point $t_2$ to be the location 
of the closest real zero of the Gram determinant $G_5$. 
Naturally, for physical reasons, this is outside of the range $[0,1]$.
In practice, we see that this causes a noticeable decrease in the number of segments that the 
adaptive approach builds, providing an important speedup.

\subsection{Numerical Study for Five-Point Massless and One-Mass Scattering}
\label{sec:NumericalStudy}

\begin{table}[tb]
    \centering
    \begin{tabular}{c|ccccc}
       weight  & 0 & 1  & 2 & 3 & 4 \\
       \# massless functions  & 1 & 10 & 80 & 493 & 473 \\
       \# one-mass functions  & 1 & 11 & 92 & 624 & 883
    \end{tabular}
    \caption{Number of linearly independent functions at each weight that contribute to the two-loop full-colour massless squared amplitude and the leading-colour one-mass squared amplitude. The corresponding number of linearly independent master integrals is $1917$ and $1383$ for the massless and one-mass case, respectively.
    }
    \label{tab:5ptMasslessNFunctions}
\end{table}

In this section, we study the behaviour of our numerical differential equation solution strategy over 
physical phase space.
We make use of a proof-of-concept double-precision \texttt{Mathematica} implementation of 
our adaptive Chebyshev approach that we provide in ancillary files that are described in \cref{sec:Ancillary}.
We consider two collections of two-loop five-point Feynman integrals. 
First, we consider the complete set of two-loop five-point massless Feynman integrals~\cite{Gehrmann:2015bfy,Gehrmann:2018yef,Abreu:2018rcw,Chicherin:2018mue,Abreu:2018aqd,Chicherin:2018old}.
Second, we consider the complete set of two-loop five-point one-mass integrals relevant for the leading 
colour production of a $W$ boson in association with two jets~\cite{Abreu:2020jxa,Canko:2020ylt}. 
We study both of these cases by using our \texttt{Mathematica} implementation to 
sample over phase space, and compare to higher precision evaluations provided by the $\texttt{C++}$ code, 
\texttt{PentagonFunctions++}~\cite{Chicherin:2020oor,Chicherin:2021dyp}.

Let us begin by describing the invariant kinematics in the channel of \cref{eq:PhysicalChannel}.
The kinematics of the complete set of two-loop five-point massless Feynman 
integrals can be specified in terms of 
five independent Mandelstam variables,
\begin{equation}
    \vec{s}^{\text{ massless}} = \{s_{12}, s_{23}, s_{34}, s_{45}, s_{15}\}. 
\end{equation}
Here, energy positivity constrains the Mandelstam variables to satisfy the inequalities
\begin{equation}
    s_{12}, s_{34}, s_{35}, s_{45} > 0, \qquad \qquad s_{13}, s_{14}, s_{15}, s_{23}, s_{24}, s_{25} < 0.
\end{equation}
For five-point one-mass scattering processes, we have that four of the external momenta are massless, while one of them is massive. We choose the massive momentum to be $p_5$ such that we have
\begin{equation}\label{eq:kin5pt1m}
     p_i^2 = 0, \quad i \, \in \, \{1, 2, 3, 4\}, \qquad  p_5^2 = Q^2.
\end{equation}
There are now six independent Mandelstam invariants given by
\begin{equation}
    \vec{s}^{\text{ 5pt1m}} = \{ s_{12}, s_{23}, s_{34}, s_{45}, s_{15}, p_5^2 \}.
    \label{eq:5pt1mMandDef}
\end{equation}
In this case, the linear constraints on the Mandelstam variables in \cref{eq:5pt1mMandDef} are given by
\begin{equation}
    s_{12}, s_{34} > 0, \qquad s_{13}, s_{14}, s_{23}, s_{24} < 0, \qquad s_{15}, s_{25} < Q^2, \qquad s_{35}, s_{45} > Q^2.
\end{equation}

To construct the system of differential equations for the pentagon functions we start from the differential equations of refs.~\cite{Abreu:2018aqd} and~\cite{Abreu:2020jxa}. We permute the system of differential equations and remove redundancies between the set of permuted Feynman integrals using \texttt{FEYNSON}~\cite{Maheria:2022dsq}.
We then use the approach of ref.~\cite{Abreu:2023rco} to construct a basis of pentagon functions $\tilde{I}_i^{(n)}$  up to weight 4.
In a nutshell, the pentagon functions are chosen to be components of master integrals, relations between functions are obtained from linear algebra operations at symbol level and then lifted to function level using a single numerical evaluation.\footnote{We note that this approach differs slightly from that of refs.~\cite{Chicherin:2020oor,Chicherin:2021dyp}, leading to a smaller number of functions. The numbers quoted in \cref{tab:5ptMasslessNFunctions} should thus not match with the ones given of refs.~\cite{Chicherin:2020oor,Chicherin:2021dyp}}
We report the number of linearly independent functions at each weight in \cref{tab:5ptMasslessNFunctions}, 
as this controls the sizes of the differential equations that we must solve.
In both cases, we used high-precision evaluations from the \texttt{PentagonFunctions++} library
as boundary conditions \cite{gitlabPentagonFunctions}.

To benchmark the performance of our differential equation solver, we test the
implementation on a MacBook Pro equipped with an \texttt{M5 Pro}
processor. We generate phase-space points in a flat manner using an internal \texttt{Mathematica}
implementation of the RAMBO algorithm~\cite{Kleiss:1985gy} such that the centre of
mass energy $s_{12}$ is normalised to 1. As an initial boundary point for the
evolution, for the massless case, we select the phase-space point given by
\begin{equation}
    \vec{s}^{\text{ massless}}_0 = \left\{\frac{29}{10}, -\frac{202}{235}, \frac{679}{745}, \frac{1239}{1285}, -\frac{2544}{2705} \right\},
\end{equation}
while for the one-mass case, we use the boundary point\footnote{\label{foot:conventions} We note that this point is not in the kinematic region implemented in the \texttt{PentagonFunctions++} code, since the conventions there are that one should work in the channel $p_4,p_5\to p_1,p_2,p_3$. It is a trivial relabeling to go from one channel to the other.}
\begin{equation}
    \vec{s}^{\text{ 5pt1m}}_0 = \left\{10, -\frac{7}{3}, \frac{9}{5}, \frac{13}{4}, -\frac{3}{2}, 1 \right\}.
\end{equation}
We note that, unlike in ref.~\cite{Chicherin:2020oor}$, \vec{s}^{\text{ massless}}_0$ is chosen such that it is not a star centre. 
As a star centre must lie on a spurious singularity surface, its use would introduce numerical complications when evaluating the integrands in \cref{eq:PathIntegrationForm}. Nevertheless, in practice, our chosen point is sufficiently well-behaved that only an insignificantly small number of phase-space points cannot be reached via a simple straight-line path.
In contrast, for the one-mass case, we find that starting at $\vec{s}^{\text{ 5pt1m}}_0$ requires us to construct a two-segment path around $8\%$ of the time.

\begin{figure}[tb]
\centering
\includegraphics[scale=1.0]{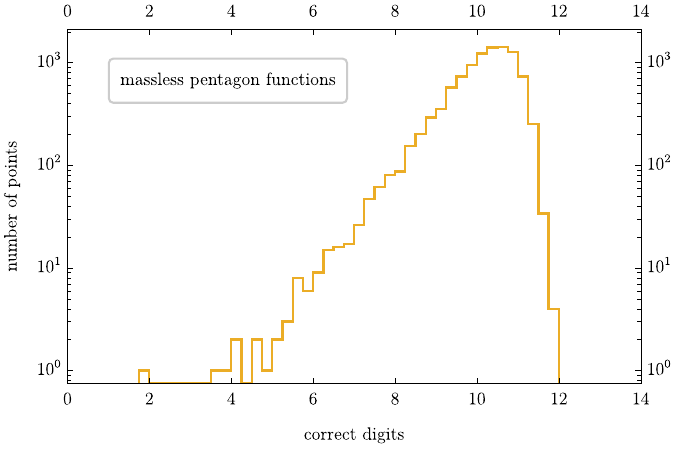}
    \caption{Precision study for five-point massless pentagon functions: distribution of correct digits,
      $-\log_{10}(E_r[\vec{s}])$, over $10^4$
      physical phase-space points as generated by RAMBO with
      $s_{12}$ normalised to 1.}
      \label{fig:masslessStability}
\end{figure}

\begin{figure}[tb]
\centering
\includegraphics[scale=1.0]{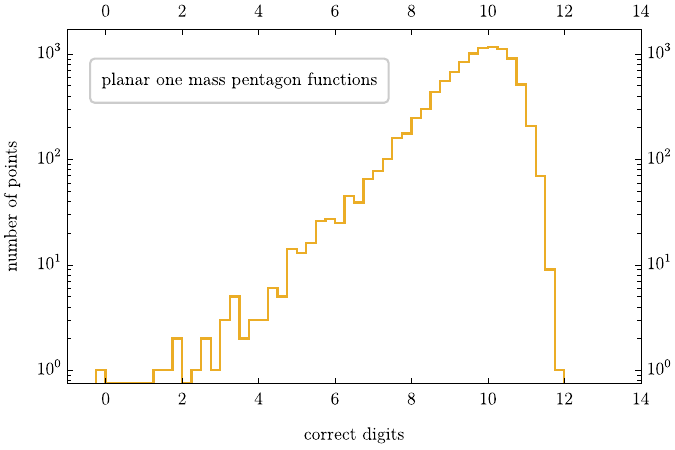}
    \caption{Precision study for five-point one-mass pentagon functions: distribution of correct digits,
      $-\log_{10}(E_r[\vec{s}])$, over $10^4$
      physical phase-space points as generated by RAMBO with
      $s_{12}$ normalised to 1.}
      \label{fig:massiveStability}
\end{figure}

To validate the accuracy of our evaluations, we compare our double-precision implementation against evaluations obtained from the \texttt{PentagonFunctions++} library in quadruple precision. Note that our framework computes a large number of special functions simultaneously, thus we report the maximum relative error observed across the entire set of functions. Specifically, we define this error measure as
\begin{equation}
    E_r(\vec{s}) = \max_{n, \,\, i} \left|1 - \frac{\tilde{I}_i^{(n), \text{cheby}}(\vec{s})}{\tilde{I}_i^{(n), \text{p++}}(\vec{s})}\right|,
    \label{eq:FunctionSetRelError}
\end{equation}
where the label ``cheby'' denotes that the function has been evaluated using the Chebyshev solver, and the label ``p++'' 
denotes that the function has been evaluated using the \texttt{PentagonFunctions++} code.
In \cref{fig:masslessStability,fig:massiveStability} we plot the error as defined 
in \cref{eq:FunctionSetRelError} across $10^4$ phase-space points generated by RAMBO for the massless and one-mass configurations, respectively. We recall that all our evaluations are performed in double precision, and in our proof-of-principle implementation, we have not implemented any rescue system to correct the points where precision is lost.
Despite this, we observe that precision is under excellent control, with catastrophic loss of precision in a vanishingly small number of points in the one-mass case. 
There are several strategies that can be explored to circumvent loss of precision within our algorithm, but we leave this for future work.

\begin{table}[t]
    \centering
    \begin{tabular}{c|c c c}
        & Median (s) & Mean (s) & Max (s) \\ 
        Massless & $0.35$ & $0.56$ & $6.5$ \\
        One-mass & $1.2$ & $1.45$ & $10.7$
    \end{tabular}
    \caption{Timing characterization of proof-of-concept, double-precision, \texttt{Mathematica} implementation of adaptive Chebyshev differential equation solver to evaluate complete sets of pentagon functions. Variation over phase space can be attributed to the use of more segments with the adaptive strategy}
    \label{tab:timings}
\end{table}

Finally, let us remark upon the performance of our implementation across phase space. We find that the evaluation is highly efficient, yielding timings that are competitive with the \texttt{PentagonFunctions++} code. We record characteristic timing evaluations in \cref{tab:timings}.
The difference in evaluation times can be attributed to a larger or smaller number of segments constructed by the algorithm. Notably, the fastest evaluations use only a single segment with $n = 16$.
Naturally, we see that the runtimes for the one-mass case are higher than the massless case, which can be attributed to the larger size of the differential equation matrices.

\subsection{Ancillary Files}
\label{sec:Ancillary}

We provide ancillary files that contain a proof-of-concept implementation 
of our adaptive Chebyshev differential-equation solver. 
The code has been used to produce the histograms in 
\cref{sec:NumericalStudy}. In order to facilitate understanding of the code, we provide example notebooks 
that can be used to evolve the two sets of pentagon functions from a boundary point to a second target point.
Our ancillary files can be found at ref.~\cite{ancfiles} and are 
organised as follows:
\begin{itemize}
    \item \texttt{/pfDefinition\_massless}: Folder containing a script that defines the pentagon functions for all five-point massless integrals up to two loops. We give our conventions for the pure basis for the full space of integrals, and a map from those integrals to our definition of pentagon functions. The latter are the functions we evaluated with our adaptive Chebyshev differential-equation solver
    \item \texttt{/pfDefinition\_1m}: The same as the folder above, now for the five-point one-mass planar integrals up to two loops. We refer back to the comment in \cref{foot:conventions}:
    the conventions of ref.~\cite{Abreu:2020jxa} and of the \texttt{PentagonFunctions++} are not directly compatible with the physical channel in \cref{eq:PhysicalChannel} and the conventions of \cref{eq:kin5pt1m}. This is, however, easily fixed by a trivial relabeling of the momenta.
    \item \texttt{source/chebyshev.m}: Proof-of-concept implementation of adaptive Chebyshev interpolation.
    \item \texttt{source/aux.wl}: Differential equation solution technology.
    \item \texttt{source/Grams.m}: Expressions for five-point Gram determinants.
    \item \texttt{DE/Alphabet\_[5ptMassless,5ptOneMass].m}: Expressions for the letters $W_k$.
    \item \texttt{DE/DE\_[5ptMassless,5ptOneMass].m}: Connection matrices $\tilde{M}_{ijk}^{(n)}$.
    \item \texttt{BC/*.m}: Benchmark boundary and target values.
    \item \texttt{[5ptMassless,5ptOneMass].wl}: Proof-of-concept notebooks demonstrating the evaluation of five-point massless and five-point one-mass pentagon functions with the Chebyshev approximation technology.
\end{itemize}

We note that the target point in the massless demonstration notebook 
was chosen to demonstrate one of the strengths of the approach, as it
lives on a spurious singularity. Specifically, we choose the point
\begin{equation}
        \vec{s}^{\text{ massless}}_1 = \left\{3,  -1, 1,  1, -1 \right\}.
\end{equation}
Evaluation on $\vec{s}^{\text{ massless}}_1$ would cause numerical issues if one were to
construct a path that ends at $\vec{s}^{\text{ massless}}_1$, as one of the Chebyshev nodes would lie on a spurious singularity. 
Instead, we construct the Chebyshev approximations on a path which contains the target point.
Once the approximation is constructed, it can be used to evaluate the solution on any
point along the path, and in particular at $\vec{s}^{\text{ massless}}_1$, providing a simple
way to sidestep the numerical issues.

Finally, we stress that the evaluation notebooks use straight-line paths and will give 
incorrect results if one 
modifies the final point such that the straight line leaves physical phase space.
As noted previously, for the phase-space study in \cref{sec:NumericalStudy}, we implemented an algorithm to combine paths when a single straight path would leave phase space.

\section{Summary and Outlook}

In this work, we have explored the application of Chebyshev approximation
techniques to the evaluation of Feynman integrals. The approach is motivated by its exponentially
convergent behaviour for analytic functions, a
characteristic that naturally complements the robust understanding of the
analyticity properties of scattering amplitudes and Feynman integrals.
We have introduced an adaptive Chebyshev approximation
framework tailored for solving canonical differential equations for linearly-independent sets of special functions.
This method dynamically detects the convergence of the approximation by
employing the plateau test of ref.~\cite{Aurentz2017}, avoiding oversampling in 
the presence of round-off errors or non-analyticities near the region of approximation.
Using this approach, we have been able to demonstrate excellent control over
precision across physical phase space.
Furthermore, we have shown that it is straightforward to construct an
implementation that relies purely on standard machine precision arithmetic.
As a practical proof of concept, we provided a \texttt{Mathematica}
implementation of two-loop five-point massless pentagon functions and one-mass planar pentagon functions
that exhibits run times which are competitive with current implementations in compiled languages~\cite{Chicherin:2020oor, Chicherin:2021dyp}.
Crucially, the approach is strongly flexible: once a canonical differential equation is
available, it can be easily and rapidly integrated within the Chebyshev
framework.

Looking forward, we see several directions to explore for future research.
A natural next step is the direct application of this framework to more
demanding kinematics, such as the collection of recently computed two-loop
six-point Feynman integrals~\cite{Henn:2021cyv, Abreu:2024fei, Henn:2024ngj,
Henn:2025xrc}.
Moreover, in order to be able to apply the technique to phenomenological
studies, it will be important to develop a public implementation optimised for
cluster-scale applications, and if possible in a compiled language that would
further improve the already excellent timings of our approach.
Furthermore, it would be highly interesting to extend this approach to the
treatment of five-point integrals with more external or even internal masses, that can involve many
roots or even elliptic sectors \cite{FebresCordero:2023pww,Abreu:2024yit,Badger:2024fgb,
Becchetti:2025oyb, Becchetti:2025qlu}.
Finally, it would be interesting to consider a systematic study of the
computation of Feynman integrals in singular regions where massless particles
become soft or collinear, for application to real/virtual contributions to
physical cross sections.

\paragraph{Note added:} While this manuscript was in the final stages of
preparation, the preprint~\cite{Liu:2026cpf} appeared, which also provides a 
public Chebyshev approximation method for Feynman integrals.

\section*{Acknowledgements}
We would like to thank S.~Devoto, M.~Kraus, X.~Liu and P.F.~Monni for helpful
discussions and collaboration on related projects.
The work of A. Guerreiro is supported by Funda\c{c}\~ao para a Ci\^encia e a Tecnologia
(FCT) under the contract 2025.03095.BD.

\bibliographystyle{JHEP}
\bibliography{biblio}
\end{document}